\documentclass[aps,prc,letterpaper,11pt,twoside,tightenlines,nofootinbib,showpacs,preprint]{revtex4-1}
\usepackage{epsfig}
\usepackage{subfigure}
\usepackage{amsmath}
\usepackage{multirow}
\begin{document}

%\linenumbers

%\bibliographystyle{unsrt}
\arraycolsep1.5pt
\newcommand{\Ima}{\textrm{Im}}
\newcommand{\Rea}{\textrm{Re}}
\newcommand{\mev}{\textrm{ MeV}}
\newcommand{\gev}{\textrm{ GeV}}

\def\bra#1#2#3#4#5{_{(\ell_M={#4}/2,\,\ell_B={#5})}\big\langle S_{c\bar
    c}={#1},\, {\cal L}=\frac{{#2}}{2}\,;  J=\frac{{#3}}{2}|}
\def\bracc#1#2#3#4{_{(\ell_M=0,\,\ell_B=\frac{{#4}}{2})}\big\langle S_{c\bar
    c}={#1},\, {\cal L}=\frac{{#2}}{2}\,; J=\frac{{#3}}{2}|}
\def\ket#1#2#3#4#5{| S_{c\bar c}={#1},\, {\cal L}=\frac{{#2}}{2}\,;J=\frac{{#3}}{2}\big\rangle_{(\ell_M={#4}/2,\,\ell_B={#5})}}
\def\ketcc#1#2#3#4{| S_{c\bar c}={#1},\, {\cal L}=\frac{{#2}}{2}\,; J=\frac{{#3}}{2}\big\rangle_{(\ell_M=0,\,\ell_B=\frac{{#4}}{2})}}

\def\ketbb#1#2#3{| S_{b\bar b}={#1},\, {\cal L}={#2}\,;\, J={#3} \,\big\rangle}
\def\ketbbs#1#2#3{| S_{b\bar b}={#1},\, {\cal L}={#2}\,;\, J={#3} \,\big\rangle_s}

\title{Hidden beauty molecules within the local hidden gauge approach and heavy quark spin symmetry.}

%\author{A. Ozpineci, C. W. Xiao$\,^1$ and E. Oset$\,^1$}

%\affiliation{
%$^1$Departamento de F\'{\i}sica Te\'orica and IFIC, Centro Mixto Universidad \\de Valencia-CSIC, Institutos de Investigaci\'on de Paterna, Apartado 22085, 46071 Valencia, Spain
%}

\author{A. Ozpineci\footnote{On sabbatical leave from Physics Department, Middle East Technical University, Ankara, Turkey}, C. W. Xiao and E. Oset}
\affiliation{
Departamento de F\'{\i}sica Te\'orica, Universidad de Valencia and
IFIC, Centro Mixto Universidad de 
Valencia-CSIC,
Institutos de Investigaci\'on de Paterna, Aptdo. 22085, 46071 Valencia,
Spain
}

\date{\today}

\begin{abstract}

Using a coupled channel unitary approach, combining the heavy quark spin symmetry and the dynamics of the local hidden gauge, we investigate the meson-meson interaction with hidden beauty and obtain several new states. Both $I=0$ and $I=1$ states are analyzed and it is shown that in the $I=1$ sector, the interactions are too weak to create any bound states within our framework. In total, we predict with confidence the existence of $6$ bound states, and weakly bound $6$ more possible states. The existence of these weakly bound states depend on the influence of the coupled channel effects.
 
\end{abstract}


\maketitle

\section{Introduction}

The world of heavy quarks, charm and beauty, is experiencing a fast development, with a plethora of new states being found in facilities as BABAR, CLEO, BELLE, BES \cite{Ali:2011vy,Gersabeck:2012rp,Olsen:2012zz,Li:2012pd}. The coming facility of FAIR will certainly add new states corresponding to quantum numbers which are not accessible with present machines. The states capturing more attention are those that do not fit within the standard picture of mesons as $q \bar q$ or baryons as $qqq$, and which require more complex structures, like tetraquarks, meson molecules, or hybrids including possible glueballs, for mesons, or pentaquarks and meson baryon molecules for baryons. 

   The field of meson molecules in the charm sector has been much studied  \cite{Kolomeitsev:2003ac,Hofmann:2003je,Guo:2006fu,dany,danyax,Branz:2009yt,Faessler:2007gv,
Segovia:2008zz,FernandezCarames:2009zz,Gutsche:2010zza,HidalgoDuque:2012pq,Guo:2008zg} and many of the observed states with hidden charm and open charm are shown to be consistent with the molecule interpretation, with a good reproduction of the different observables of those states. The work on the charm sector is gradually moving to the beauty sector and there are works dealing with $b$ or hidden $b$ meson molecules built up from other mesons containing some $b$ quark \cite{Voloshin:1982ij,Ding:2009vj,Bondar:2011ev,juanmanolo,Cleven:2011gp,Li:2012ss,hanhart,Li:2012mqa}. The recent discovery of the hidden beauty $Z_b(10610)$ and $Z_b(10650)$ states in three charge states \cite{Belle:2011aa,Adachi:2012cx,Adachi:2012im}, and hence with isospin I=1, has brought a new stimulus to the molecular idea \cite{juanmanolo,Cleven:2011gp,hanhart}, since they cannot be  $b \bar b$ quarkonium states. 
   
   One of the elements that has allowed progress in the heavy quark sector is the implementation of the heavy quark spin symmetry (HQSS) \cite{Isgur:1989vq,Neubert:1993mb,manohar,guosym}. QCD predicts that all types of spin interactions vanish for infinitely massive quarks and thus, for heavy quarks the dynamics is unchanged under arbitrary transformations of their spin. This heavy quark spin independence is the essence of the HQSS, and leads to many predictions concerning properties of particles containing heavy quarks. It also implies that the leading interaction terms are independent of the flavor. However, the HQSS does not determine the interaction, simply puts some constraints in it, and to proceed further to make predictions one must rely upon some experimental information or use models. In this sense, the work of \cite{juanmanolo} uses properties from the X(3872) resonance, which is assumed to be a $D \bar D^* -cc$ molecule, and extrapolates this information to make predictions of $B \bar B^* -cc$ molecules. 
   
   An alternative approach to using empirical data to constraint the interaction  is the use of some dynamical model. The use of chiral Lagrangians has been a common thing in this kind of works, but its extension to the heavy quark sector is not straightforward. Conversely, the use of the local hidden gauge Lagrangians has allowed much progress in the heavy sector. Indeed, the Lagrangians of this theory  \cite{hidden1,hidden2,hidden4}, introducing pseudoscalar and vector mesons as building blocks, provides the same information as the chiral Lagrangians up to next to leading order under the assumption of vector meson dominance \cite{sakurai}, and additionally introduces explicitly vector mesons and their interaction in the theory.  This feature is much welcome when working in the heavy sector, because the independence on the spin of the heavy quarks puts at the same level the $D$ and $D^*$ and the $B$ and $B^*$, and one has to deal with them simultaneously. The other bonus from using the hidden gauge formalism is that it shows that the leading chiral Lagrangian is obtained by exchanging vector mesons between the interacting particles. In the SU(3) sector, these are the $\rho,\omega,\phi, K^*$.  In the charm or beauty sector the $D$ of $B$ mesons contain a light quark, and in their interaction it is these light quarks that are exchanged with vector-meson quantum numbers, and then the analogy with the interaction in the light sector becomes apparent. Of course in addition one might also exchange heavy vectors in the charm or beauty sectors, but their large mass renders their propagators small and those terms become subdominant.  From this perspective it looks quite intuitive that many results obtained in the SU(3) sector can be extrapolated to the heavy sector. What is less intuitive, but can be seen from the dynamics, is the recent finding in \cite{juanxiao} that the local hidden gauge dynamics fully respects the constrains of HQSS. In this latter work, baryons with hidden charm are investigated. The resulting theoretical framework is also extended to study baryon states of hidden beauty in \cite{xiaooset}. 
   
   The finding that the local hidden gauge approach respects HQSS is quite relevant since it allows one to be more predictive, even if some phenomenology is still needed to regularize the loops of the theory. But certainly, having the interaction given by the hidden gauge Lagrangians allows one to tackle theoretically many problems where there is scarce or no phenomenology, as is the case of the beauty sector. 
   
   This is the purpose of the present paper where we tackle the interaction of $B \bar B$, $B \bar B^*$ and $B^* \bar B^*$ in the hidden beauty sector and make predictions for bound states.  We will find bound states in all sectors and in sum we get six states with different quantum numbers. The states have their correspondence with states already seen, or predicted in the hidden charm sector. One bound $B \bar B$ state would correspond to the so called X(3700), predicted in \cite{dany} for which experimental support is obtained from the 
$e^+ e^- \to J / \psi D \bar D$ reaction in \cite{Gamermann:2007mu}. Two bound states are also obtained for $B \bar B^*$ in analogy to the $D \bar D^*$ X(3872) resonance that has C-parity positive, plus an extra one with negative C-parity, and three more states are obtained from $B^* \bar B^*$ corresponding to states with different spins obtained for the $D^* \bar D^*$ system in \cite{raquelxyz}. All the states obtained have isospin I=0. By using the leading term from the HQSS generated from the local hidden gauge approach, we do not get I=1 states. Thus, the interesting $Z_b(10610)$ and $Z_b(10650)$ states are not generated at this level. The fact that these states are weakly unbound from the 
$B \bar B^*$ and $B^* \bar B^*$ thresholds could hint at subdominant terms in HQSS responsible for their generation, thus justifying why they do not appear just using the dominant term.

\section{HQSS Formalism}

Following the work of \cite{juanxiao} for hidden charm baryons, we extrapolate the formalism to the hidden beauty sector for the mesons. Therefore we can study mesons with hidden beauty with isospin $I=0,\ 1$, and spin $J=0,\ 1,\ 2$. We take as coupled channels states with $B,\ B^*$, $B_s,\ B^*_s$ and their corresponding antiparticles. For the different $I,\ J$ quantum numbers we have the following space states.\\

1) $J=0,\ I=0$

$\quad B \bar{B},\ B_s \bar{B}_s,\ B^* \bar{B}^*,\ B^*_s \bar{B}^*_s.$\\

2) $J=0,\ I=1$

$\quad B \bar{B},\ B^* \bar{B}^*.$\\

3) $J=1,\ I=0$

$\quad B \bar{B}^* ~(B^* \bar{B}),\ B_s \bar{B}_s^* ~(B_s^* \bar{B}_s), \ B^* \bar{B}^*,\ B^*_s \bar{B}^*_s$.\\

4) $J=1,\ I=1$

$\quad B \bar{B}^* ~(B^* \bar{B}), \ B^* \bar{B}^*$.\\

5) $J=2,\ I=0$

$\quad B^* \bar{B}^*,\ B^*_s \bar{B}^*_s$.\\

6) $J=2,\ I=1$

$\quad B^* \bar{B}^*$.\\

With different spin quantum number there are 12 orthogonal states (of which 6 are having hidden strangeness) in the physical basis for $I=0$. For $I=1$ there are only 6 states since the hidden strangeness states have $I=0$. Next we will introduce a HQSS basis \cite{juanxiao}, for which it is straightforward to implement the lowest order HQSS constraints. In the HQSS basis we will classify the states in terms of the quantum numbers, $J$: total spin of the meson-baryon system, ${\cal L}$: total spin of the light quarks system, $S_{b \bar{b}}$: total spin of the  $b \bar{b}$ subsystem. Note that we study ground state mesons, which means that all orbital angular momenta are zero.

Thus, we have 12 orthogonal states in the physical basis. The 12 orthogonal states in the HQSS basis are given by
\begin{itemize}
\item {\small $\ketbb 000$, $\ketbbs 000$,}
\item {\small $\ketbb 011$, $\ketbbs 011$,}
\item {\small $\ketbb 101$, $\ketbbs 101$,}
\item {\small $\ketbb 110$, $\ketbbs 110$,}
\item {\small $\ketbb 111$, $\ketbbs 111$,}
\item {\small $\ketbb 112$, $\ketbbs 112$.}
\end{itemize}
The subindex $s$ in the former states means that the light quarks are strange.

In order to take into account the HQSS it is interesting to use the heavy quark basis in which the spins are rearranged such as to 
combine the spin of the $b \bar{b}$ quarks into $S_{b \bar{b}}$ since the matrix elements do not depend on this spin.
 One classifies the HQSS in terms of $\vec{S}_{b \bar{b}}, ~\vec{\cal L}$ and $\vec{J}$. 
 The conservation of $\vec{S}_{b \bar{b}}$ and $\vec{J}$ leads to the conservation of $\vec{\cal L} = \vec{J} - \vec{S}_{b \bar{b}}$ and then in the HQSS basis the matrix elements fulfil
\begin{equation}
\big\langle S'_{b\bar b},\, {\cal L}'; J',\, \alpha'|H^{QCD}|
S_{b\bar b},\, {\cal L}; J,\, \alpha \big \rangle 
=  \;
\delta_{\alpha \alpha'}\delta_{JJ'}\delta_{S'_{b\bar b}S_{b\bar b} }
\delta_{{\cal L}{\cal L}'}  \big\langle {\cal L};
\alpha ||H^{QCD}  || {\cal L}; \alpha \big\rangle. \label{eq:hqs}
\end{equation}

 Thus, in a given $\alpha$ sector, we have a total of six unknown low energy constants (LEC's):
\begin{itemize}
\item Three LEC's associated to ${\cal L}=0$
\begin{eqnarray}
\lambda_0^\alpha &=& \big\langle {\cal L}=0;\alpha ||H^{QCD}  || {\cal L}=0;\alpha
 \big\rangle \\
\lambda_{0s}^\alpha &=& _s \big\langle {\cal L}=0;\alpha ||H^{QCD}  || {\cal L}=0;\alpha
 \big\rangle_s \\
\lambda_{0m}^\alpha &=& \big\langle {\cal L}=0;\alpha ||H^{QCD}  || {\cal L}=0;\alpha
 \big\rangle_s
\end{eqnarray}
\item Three LEC's associated to ${\cal L}=1$
\begin{eqnarray}
\lambda_1^\alpha &=& \big\langle {\cal L}=1;\alpha ||H^{QCD}  || {\cal L}=1;\alpha
 \big\rangle \\
\lambda_{1s}^\alpha &=& _s \big\langle {\cal L}=1;\alpha ||H^{QCD}  || {\cal L}=1;\alpha
 \big\rangle_s \\
\lambda_{1m}^\alpha &=& \big\langle {\cal L}=1;\alpha ||H^{QCD}  || {\cal L}=1;\alpha
 \big\rangle_s \label{eq:lambda}
\end{eqnarray}
\end{itemize}
Therefore in the HQSS basis, the  $H^{QCD}$ is a block diagonal matrix.

To exploit Eq.~(\ref{eq:hqs}), one should express hidden beauty uncoupled meson-meson states in terms of the HQSS basis.Therefore, one needs to use 9-j symbols.

The 9-j symbols are used to relate two basis where the angular momentums are coupled in a different way. Taking two particles with $\vec{l}_1,\ \vec{s}_1$ and $\vec{l}_2,\ \vec{s}_2$, we can combine them to $\vec{j}_1,\ \vec{j}_2$ and finally $\vec{j}_1,\ \vec{j}_2$ to total $\vec{J}$. Alternatively we can couple $\vec{l}_1,\ \vec{l}_2$ to $\vec{L}$, $\vec{s}_1,\ \vec{s}_2$ to $\vec{S}$, and then $\vec{L},\ \vec{S}$ to total $\vec{J}$. These two bases are related as \cite{Rose}
\begin{equation}
\begin{split}
|l_1 s_1 j_1; l_2 s_2 j_2; J M \big\rangle =& \sum_{S,L} [ (2 S + 1) (2 L + 1) (2 j_1 + 1) (2 j_2 + 1)]^{1/2} \\
& \times \left\{
\begin{array}{ccc}
l_1 & l_2 & L \\
s_1 & s_2 & S \\
j_1 & j_2 & J
\end{array}
\right\} \ |l_1 l_2 L; s_1 s_2 S; J M \big\rangle,
\end{split}
\end{equation}
where the symbol $\{ \}$ stands for the 9-j coefficients.

\begin{figure}[tb]
\epsfig{file=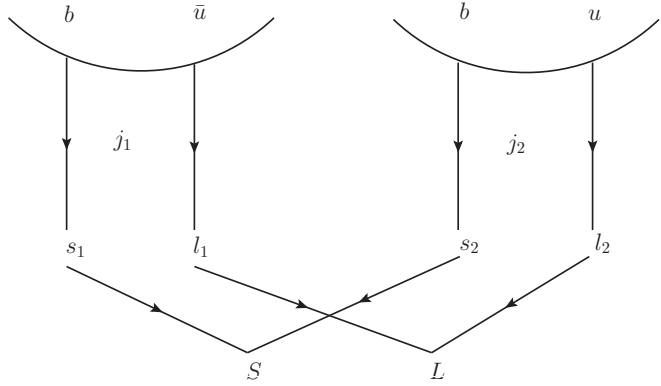, width=9cm}
\caption{Diagrams for the calculation of 9-j coefficients.}%
\label{fig:9j}%
\end{figure}

As an example take a meson(M)-antimeson($\bar{M}$) state of the type $B^{(*)} \bar{B}^{(*)}$ and look at the recombination scheme on Fig. \ref{fig:9j}.
Thus in this case we have the correspondence,
\begin{align*}
&\text{generic:} && l_1 && l_2 && s_1 && s_2 && j_1 && j_2 && L && S && J  \\
&\text{HQSS:} && \frac{1}{2} && \frac{1}{2} && \frac{1}{2} && \frac{1}{2} && J_M(0,1) && J_{\bar M}(0,1) && {\cal L}(0,1) && S_{b \bar{b}}(0,1) && J (0,1,2) \ .
\end{align*}
%\omega
with $J_M$ and $J_{\bar M}$ the total spin of the meson and antimeson respectively. Then one easily finds:
\begin{itemize}
\item $J=0$
\begin{eqnarray}
|B \bar B \big \rangle &=& \frac12 \ketbb000 + \frac{\sqrt{3}}{2} \ketbb110 
\label{hqssbasisfirst}\\
\nonumber \\
|B^* \bar B^* \big \rangle &=& -\Big(\frac{\sqrt{3}}{2} \ketbb000 - \frac12 \ketbb110\Big) \\
\nonumber \\
|B_s \bar B_s \big \rangle &=& \frac12 \ketbbs000 + \frac{\sqrt{3}}{2} \ketbbs110 \\
\nonumber \\
|B^*_s \bar B^*_s \big \rangle &=& -\Big(\frac{\sqrt{3}}{2} \ketbbs000 - \frac12 \ketbbs110\Big)
\end{eqnarray}

\item $J=1$
\begin{eqnarray}
|B \bar B^* \big \rangle &=& -\Big(-\frac12 \ketbb101 + \frac12  \ketbb011 \nonumber \\ 
&+& \frac{\sqrt{2}}2 \ketbb111\Big) \\
\nonumber \\
|B^* \bar B \big \rangle &=& \frac12 \ketbb101 - \frac12  \ketbb011 \nonumber \\ 
&+& \frac{\sqrt{2}}2 \ketbb111 \\
\nonumber \\
|B^* \bar B^* \big \rangle &=& -\Big(\frac{\sqrt{2}}2 \ketbb101 + \frac{\sqrt{2}}2 \ketbb011\Big) \\
\nonumber \\
|B_s \bar B^*_s \big \rangle &=& -\Big(-\frac12 \ketbbs101 + \frac12  \ketbbs011 \nonumber \\ 
&+& \frac{\sqrt{2}}2 \ketbbs111\Big) \\
\nonumber \\
|B^*_s \bar B_s \big \rangle &=& \frac12 \ketbbs101 - \frac12  \ketbbs011 \nonumber \\ 
&+& \frac{\sqrt{2}}2 \ketbbs111 \\
\nonumber \\
|B^*_s \bar B^*_s \big \rangle &=& -\Big(\frac{\sqrt{2}}2 \ketbbs101 + \frac{\sqrt{2}}2 \ketbbs011 \Big)
\end{eqnarray}

\item $J=2$
\begin{eqnarray}
|B^* \bar B^* \big \rangle &=& -\ketbb112 \\
\nonumber \\
|B^*_s \bar B^*_s \big \rangle &=& -\ketbbs112
\end{eqnarray}

\end{itemize}

All the states with $\bar B^*$ (or $\bar B^*_s$) have an extra minus sign in the former formulas. This is because we should take into account the C-parity for $B \bar B^*$ and $B^* \bar B$ ($B_s \bar B^*_s$ and $B^*_s \bar B_s$), as discussed in \cite{HidalgoDuque:2012pq}. We know that $\hat{C} \rho^0 = - \rho^0$, and then, we take $\hat{C} B^* = - \bar B^*$, thus, $\bar B^* = -\hat{C} B^*$. Then, the $\bar B^*$ behaves like $B^*$ by changing $q \to \bar q$ but with a minus sign. Therefore, we should put a minus sign to all the states which are constructed with one $\bar B^*$ ($\bar B^*_s$) particle, listed above with HQSS basis. One can easily check that the constructed state $(B \bar B^* - B^* \bar B)/\sqrt{2}$ has positive C-parity, and $(B \bar B^* + B^* \bar B)/\sqrt{2}$ negative C-parity (the ones with hidden strangeness are similar). Then, we have
\begin{eqnarray}
|B \bar B^* (B^* \bar B), \,C=+ \big \rangle &=& -\ketbb111 \\
\nonumber \\
|B \bar B^* (B^* \bar B), \,C=- \big \rangle &=& \frac1{\sqrt{2}} \big(\ketbb101 -  \ketbb011\big) \\ 
\nonumber \\
|B_s \bar B^*_s (B^*_s \bar B_s), \,C=+ \big \rangle &=& -\ketbbs111 \\
\nonumber \\
|B_s \bar B^*_s (B^*_s \bar B_s), \,C=- \big \rangle &=& \frac1{\sqrt{2}} \big(\ketbbs101 -  \ketbbs011\big)
\label{hqssbasislast}
\end{eqnarray}

Keeping in mind these changes required by the C-parity of the constructed states, now we can evaluate the transition matrix elements between the physical states. By taking into account the former states in the HQSS basis, and using Eqs. \eqref{eq:hqs}$-$\eqref{eq:lambda} we obtain the transition matrix elements in the physical basis.
\begin{itemize}
\item $J=0$, $I=0$
\[
\left. \phantom{(}
\begin{array}{cccc}
\phantom{-\frac{\sqrt{3}}4 \text{$\lambda_{0m}$}+\frac{\sqrt{3}}4  \text{$\lambda_{1m}$}} &
\phantom{-\frac{\sqrt{3}}4 \text{$\lambda_{0m}$}+\frac{\sqrt{3}}4  \text{$\lambda_{1m}$}} &
\phantom{-\frac{\sqrt{3}}4 \text{$\lambda_{0m}$}+\frac{\sqrt{3}}4  \text{$\lambda_{1m}$}} &
\phantom{-\frac{\sqrt{3}}4 \text{$\lambda_{0m}$}+\frac{\sqrt{3}}4  \text{$\lambda_{1m}$}} \\
B \bar{B} & B^* \bar{B}^* &  B_s \bar{B}_s &  B^*_s \bar{B}^*_s 
\end{array}
\right. \phantom{)_{I=0}}
\]
\begin{equation}
\left(
\begin{array}{cccc}
 \frac14 \text{$\lambda_0$}+\frac34  \text{$\lambda_1$} &
 -\frac{\sqrt{3}}4 \text{$\lambda_0$}+\frac{\sqrt{3}}4  \text{$\lambda_1$} &
  \frac14 \text{$\lambda_{0m}$}+\frac34  \text{$\lambda_{1m}$} &
   -\frac{\sqrt{3}}4 \text{$\lambda_{0m}$}+\frac{\sqrt{3}}4  \text{$\lambda_{1m}$}  \\ \\
 -\frac{\sqrt{3}}4 \text{$\lambda_0$}+\frac{\sqrt{3}}4  \text{$\lambda_1$} &
  \frac34 \text{$\lambda_0$}+\frac14  \text{$\lambda_1$} &
   -\frac{\sqrt{3}}4 \text{$\lambda_{0m}$}+\frac{\sqrt{3}}4  \text{$\lambda_{1m}$} &
    \frac34 \text{$\lambda_{0m}$}+\frac14  \text{$\lambda_{1m}$}  \\ \\
 \frac14 \text{$\lambda_{0m}$}+\frac34  \text{$\lambda_{1m}$} &
  -\frac{\sqrt{3}}4 \text{$\lambda_{0m}$}+\frac{\sqrt{3}}4  \text{$\lambda_{1m}$} &
  \frac14 \text{$\lambda_{0s}$}+\frac34  \text{$\lambda_{1s}$}  &
   -\frac{\sqrt{3}}4 \text{$\lambda_{0s}$}+\frac{\sqrt{3}}4  \text{$\lambda_{1s}$}  \\ \\
 -\frac{\sqrt{3}}4 \text{$\lambda_{0m}$}+\frac{\sqrt{3}}4  \text{$\lambda_{1m}$} &
  \frac34 \text{$\lambda_{0m}$}+\frac14  \text{$\lambda_{1m}$} &
   -\frac{\sqrt{3}}4 \text{$\lambda_{0s}$}+\frac{\sqrt{3}}4  \text{$\lambda_{1s}$} &
    \frac34 \text{$\lambda_{0s}$}+\frac14  \text{$\lambda_{1s}$}  \\ \\
\end{array}
\right)_{ I=0}
\label{eq:ji00}
\end{equation}
\newpage

\item $J=1(C=-)$, $I=0$
\[
\left. \phantom{(}
\begin{array}{cccc}
\phantom{\frac12 (-\text{$\lambda_{0m}$} + \text{$\lambda_{1m}$})} &
\phantom{\frac12 (-\text{$\lambda_{0m}$} + \text{$\lambda_{1m}$})} &
\phantom{\frac12 (-\text{$\lambda_{0m}$} + \text{$\lambda_{1m}$})} &
\phantom{\frac12 (-\text{$\lambda_{0m}$} + \text{$\lambda_{1m}$})} \\
B \bar{B}^* & B^* \bar{B}^* &  B_s \bar{B}_s^* &  B^*_s \bar{B}^*_s 
\end{array}
\right. \phantom{)_{I=0}}
\]
\begin{equation}
\left(
\begin{array}{cccccc}
 \frac12 (\text{$\lambda_0$} + \text{$\lambda_1$})&
 \frac12 (-\text{$\lambda_0$} + \text{$\lambda_1$}) &
  \frac12 (\text{$\lambda_{0m}$} + \text{$\lambda_{1m}$}) &
   \frac12 (-\text{$\lambda_{0m}$} + \text{$\lambda_{1m}$})  \\ \\
 \frac12 (-\text{$\lambda_0$} + \text{$\lambda_1$}) &
  \frac12 (\text{$\lambda_0$} + \text{$\lambda_1$}) &
   \frac12 (-\text{$\lambda_{0m}$} + \text{$\lambda_{1m}$}) &
    \frac12 (\text{$\lambda_{0m}$} + \text{$\lambda_{1m}$})  \\ \\
 \frac12 (\text{$\lambda_{0m}$} + \text{$\lambda_{1m}$}) &
  \frac12 (-\text{$\lambda_{0m}$} + \text{$\lambda_{1m}$}) &
  \frac12 (\text{$\lambda_{0s}$} + \text{$\lambda_{1s}$})  &
   \frac12 (-\text{$\lambda_{0s}$} + \text{$\lambda_{1s}$})  \\ \\
 \frac12 (-\text{$\lambda_{0m}$} + \text{$\lambda_{1m}$}) &
  \frac12 (\text{$\lambda_{0m}$} + \text{$\lambda_{1m}$}) &
   \frac12 (-\text{$\lambda_{0s}$} + \text{$\lambda_{1s}$}) &
    \frac12 (\text{$\lambda_{0s}$} + \text{$\lambda_{1s}$})  \\ \\
\end{array}
\right)_{ I=0}
\label{eq:ji101}
\end{equation}

%\newpage
\item $J=1(C=+)$, $I=0$
\[
\left. \phantom{(}
\begin{array}{cc}
\phantom{\text {$\lambda_{1m}$ } } &
\phantom{\text {$\lambda_{1m}$ } } \\
B \bar{B}^* &  B_s \bar{B}_s^* 
\end{array}
\right. \phantom{)_{I=0}}
\]
\begin{equation}
\left(
\begin{array}{cc}
  \text{ $\lambda_{1}$ } &
   \text{ $\lambda_{1m}$ } \\ \\
 \text {$\lambda_{1m}$ } &
  \text{ $\lambda_{1s}$ }  \\
\end{array}
\right)_{ I=0}
\label{eq:ji102}
\end{equation}

\item  $J=2$, $I=0$
\[
\left. \phantom{(}
\begin{array}{cc}
\phantom{\text {$\lambda_{1m}$ } } &
\phantom{\text {$\lambda_{1m}$ } } \\
B^* \bar{B}^* &  B_s^* \bar{B}_s^* 
\end{array}
\right. \phantom{)_{I=0}}
\]
\begin{equation}
\left(
\begin{array}{cc}
  \text{ $\lambda_{1}$ } &
   \text{ $\lambda_{1m}$ } \\ \\
 \text {$\lambda_{1m}$ } &
  \text{ $\lambda_{1s}$ }  \\
\end{array}
\right)_{ I=0}
\label{eq:ji20}
\end{equation}

\end{itemize}
For $I=1$ one removes the $B_s, ~B_s^*$ states in the former Eqs. \eqref{eq:ji00}$-$\eqref{eq:ji20}. The coefficients $\lambda_{i}^I$, $\lambda_{is}^I$ and $\lambda_{im}^I$ ($i=0,1$) are the six unknown LEC's of HQSS, which depend on isospin and can be related using $SU(3)$ flavour symmetry. The values of these coefficients are also dependent on the model used. Following Refs. \cite{juanxiao,xiaooset} we also determine them in the next section by using the local hidden gauge approach.

\section{Calculation of the LEC's with the local hidden gauge formalism}

\begin{figure}[tb]
\epsfig{file=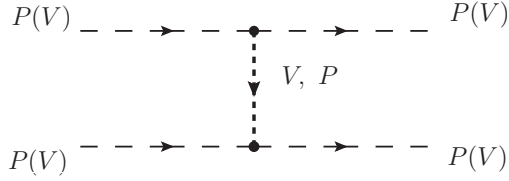, width=7cm}% \epsfig{file=f1b.eps, width=7cm}
\caption{Diagrams for interaction of pseudoscalar or vector mesons with themselves by means of meson exchange.}\label{fig:hidden}
\end{figure}

In the formalism of the local hidden gauge \cite{hidden1,hidden2,hidden4}, the Lagrangians involving the exchanged vector mesons are given by
\begin{eqnarray}
{\cal L}_{VVV} &=& ig ~\langle [V_\nu,\partial_{\mu}V_\nu]V^{\mu}\rangle, \label{eq:vvv} \\
{\cal L}_{VPP} &=& -ig ~\langle [P,\partial_{\mu}P]V^{\mu}\rangle, \label{eq:vpp} \\
{\cal L}_{PVV} &=& \frac{G'}{\sqrt{2}} ~\epsilon^{\mu \nu \alpha \beta} ~\langle \partial_{\mu} V_{\nu}  \partial_{\alpha} V_{\beta} P \rangle, \label{eq:pvv} 
\end{eqnarray}
where $g=m_V/2f$ with $f=93$~MeV the pion decay constant and taking $m_V = m_\rho$, and $G' = 3 m^2_V / (16 \pi^2 f^3)$ \cite{Bramon:1992kr,Aceti:2012cb}. The field $V_\mu$ is the SU(4) matrix of the vectors of the meson 15-plet + singlet, $P$ the SU(4) matrix of the pseudoscalar fields \cite{wuprl,wuprc}. Starting from these Lagrangians, the $PP \to PP$, $PV \to PV$, $VV \to VV$, $PP \to VV$ and $PV \to VV$ interactions can be obtained using the Feynman diagrams by exchanging a vector meson or a pseudoscalar depending on the case, as depicted in Fig. \ref{fig:hidden}.

We have formally used $SU(4)$ to obtain the vertices that are used to calculate the LECs. This symmetry is broken to $SU(3)$ by the large mass of the mesons containing the heavy $b$ quark. In the heavy quark limit (for the $b$ quark), the mesons that contain the heavy quark are not dynamic, i.e. they can not be exchanged as a virtual particle.
In the finite (but large) mass limit, these terms are suppressed by $(m_{B^*})^{-2}$ and they will be ignored in the present work. The remaining diagrams contain only virtual vector mesons made of light quarks. The couplings of these
mesons to the heavy mesons are governed by the remaining $SU(3)$ symmetry.

%We have formally used $SU(4)$ to obtain the vertices that we use to calculate the LECs.  Actually for the leading terms in the HQSS limit, we do not need to invoke $SU(4)$. The same matrix elements are obtained with $SU(3)$ substituting the $s$ quark by the $b$ quark. This can be easily understood since in the local hidden gauge approach, only light quarks are exchanged in the exchanged vectors and the heavy quarks are spectators. The dynamics of $SU(4)$ would enter in the exchange of heavy vectors, but there terms are largely suppressed since they are roughly proportional to $(m_{B^*})^{-2}$, and they are ignored in the present work.

\begin{figure}[tb]
\epsfig{file=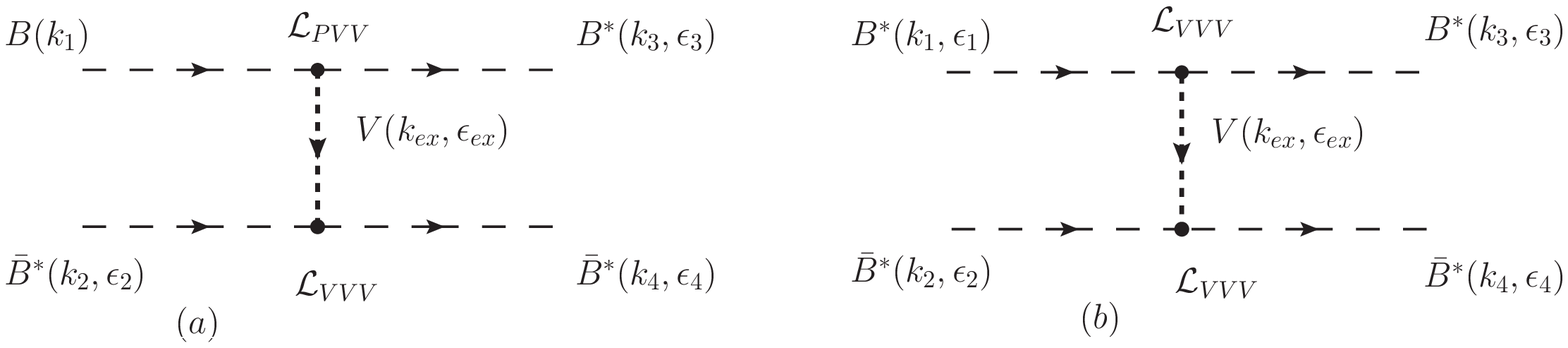, width=15cm}% \epsfig{file=f1b.eps, width=7cm}
\caption{Diagrams for interactions of $B \bar{B}^* \to B^* \bar{B}^*$ and $B^* \bar{B}^* \to B^* \bar{B}^*$.}\label{fig:hidden2}
\end{figure}

With lowest order HQSS constraints, the six unknown LEC's of $\lambda_{i}^I$, $\lambda_{is}^I$ and $\lambda_{im}^I$ ($i=0,1$) are spin independent. Therefore, we can determine them with the hidden gauge approach by some selected channels, taking the transitions $B \bar{B}^* \to B^* \bar{B}^*$ and $B^* \bar{B}^* \to B^* \bar{B}^*$ for example, shown in Fig. \ref{fig:hidden2}.
In the upper vertex of Fig. \ref{fig:hidden2} (a), using Eq. \eqref{eq:pvv}, we can have
\begin{equation}
t_{PVV} \simeq \epsilon^{\mu \nu \alpha \beta} \, k_{3\mu} \, \epsilon_{3\nu} \, k^{ex}_{\alpha} \, \epsilon^{ex}_{\beta}. \label{eq:tpvv}
\end{equation}
For the lower vertex of Fig. \ref{fig:hidden2} (a), using Eq. \eqref{eq:vvv}, we obtain
\begin{equation}
t_{VVV} = \frac{g}{\sqrt{2}} \, (k_2 + k_4)_{\mu} \, \epsilon_{2\nu} \, \epsilon_4^{\nu} \, \epsilon_{ex}^{\mu}.
\end{equation}
Thus, from these results, we can estimate the magnitude about the amplitude of $B \bar{B}^* \to B^* \bar{B}^*$. Working relatively close to threshold, as is our case, the external momenta are small, which means that $\vec{k_3} \approx 0$ and then only the $\mu = 0$ component of Eq. \eqref{eq:tpvv} contributes, $k_3^0 \approx m_{B^*}$. Thus,
\begin{equation}
t_{PVV} \sim \epsilon^{ijk} \, m_{B^*} \, \epsilon_{3i} \, k^{ex}_j \, \epsilon^{ex}_k,
\end{equation}
which implies that the momentum of the exchange vector is only spatial. Next, for the transition of $B \bar{B}^* \to B^* \bar{B}^*$, we can find
\begin{equation}
t_{B \bar{B}^* \to B^* \bar{B}^*} \sim \epsilon^{ijk} \, m_{B^*} \, \epsilon_{3i} \, k^{ex}_j \, (k_2 + k_4)_{k} \, \epsilon_{2\nu} \, \epsilon_4^{\nu} \sim k_l^2 \, m_{B^*}, \label{eq:tbbb}
\end{equation}
where $k_l^2$ is the magnitude of an external three momenta and is small. On the other hand, for the Fig. \ref{fig:hidden2} (b), the transition of $B^* \bar{B}^* \to B^* \bar{B}^*$, we can easily get
\begin{equation}
t_{B^* \bar{B}^* \to B^* \bar{B}^*} \simeq (k_1 + k_3) \cdot (k_2 + k_4)\, \epsilon_{1\mu} \, \epsilon_3^{\mu} \, \epsilon_{2\nu} \, \epsilon_4^{\nu} \sim 4 m_{B^*}^2, \label{eq:tbsbsb}
\end{equation}
where we take an approximation of $k_i^0 \approx m_{B^*} ~(i=1,2,3,4)$. Comparing with Eqs. \eqref{eq:tbbb} and \eqref{eq:tbsbsb}, we can conclude that the contribution of the transition of Fig. \ref{fig:hidden2} (a) is anomalous and subleading and then, we can take $t_{B \bar{B}^* \to B^* \bar{B}^*} \approx 0$, which is indeed the case if the actual evaluation is done. In the leading order in the $m_Q$ counting where $B$ and $B^*$ have the same mass, the term would go to zero. Therefore, we obtain
\begin{equation}
\begin{split}
&t_{B \bar{B}^* \to B^* \bar{B}^*} = \frac12 (-\lambda_0 + \lambda_1) \approx 0, \\
&\Longrightarrow \quad \lambda_0 = \lambda_1. \label{eq:degeneracy1}
\end{split}
\end{equation}
Analogously, we have
\begin{eqnarray}
\lambda_{0s} &=& \lambda_{1s}, \\
\lambda_{0m} &=& \lambda_{1m}.
\label{eq:degeneracy}
\end{eqnarray}
So, one can see that some non diagonal elements of Eqs. \eqref{eq:ji00}$-$\eqref{eq:ji20} are zero in our hidden gauge model. We can repeat the same arguments in cases where the exchange of a pseudoscalar is allowed, like $VV \to PP$. Once again, one has the three momentum squared of the exchanged pseudoscalar and the term is again subdominant. An important consequence of this, is that in our approach the $B$ and $B^*$ do not mix and then we get states for $B \bar{B}$, $B \bar{B}^*$ and $B^* \bar{B}^*$ independently, but coupled to the corresponding channels with $B_s, ~B_s^*$.

Since the LEC's are dependent on the isospin, we should take into account of the isospin structure of the states. The isospin multiplets of $B, ~B^*$ are like for kaons and we have the isospin doublets $\{ B^+,\, B^0 \}$, $\{ \bar B^0,\, -B^- \}$ and the same for $B^*$. Hence,
\begin{eqnarray}
|B^* \bar{B}^* \rangle^{I=0} &=& -\frac{1}{\sqrt{2}} (|B^{*+} \bar{B}^{*-} \rangle + |B^{*0} \bar{B}^{*0} \rangle), \\
|B^* \bar{B}^* \rangle^{I=1} &=& -\frac{1}{\sqrt{2}} (|B^{*+} \bar{B}^{*-} \rangle - |B^{*0} \bar{B}^{*0} \rangle).
\end{eqnarray}
We can easily derive $\lambda_1^{I=0}, ~\lambda_1^{I=1}$ for $B^* \bar{B}^* \to B^* \bar{B}^*$ by exchanging $\rho, ~\omega$ using Eq. \eqref{eq:vvv}, ignoring possible terms with $\Upsilon$ exchange which are negligible, and we find
\begin{eqnarray}
\lambda_1^{I=0} &=& t_{B^* \bar{B}^* \to B^* \bar{B}^*}^{I=0} = \frac14 \, g^2 (\frac{3}{m^2_{\rho}} + \frac{1}{m^2_{\omega}})(4 m^2_{B^*} - 3s), \label{eq:lam1i0} \\
\lambda_1^{I=1} &=& t_{B^* \bar{B}^* \to B^* \bar{B}^*}^{I=1} = \frac14 \, g^2 (-\frac{1}{m^2_{\rho}} + \frac{1}{m^2_{\omega}})(4 m^2_{B^*} - 3s). \label{eq:lam1i1}
\end{eqnarray}
We are also neglecting here the contact terms of the vector vector interactions of the hidden gauge approach. These terms are of order $(m_V/m_{B^*})^2$ with respect to the vector meson exchange terms \cite{raquelxyz}, and hence negligible. In the charm sector they are small but not negligible (of the order of 20 \%) and they are kept in \cite{raquelxyz}.

By taking $m_{\rho} \approx m_{\omega} = m_V$ in Eqs. \eqref{eq:lam1i0} and \eqref{eq:lam1i1}, we get a general result,
\begin{eqnarray}
\lambda_1^{I=0} &=&  \frac14 \, g^2 (\frac{3}{m^2_{\rho}} + \frac{1}{m^2_{\omega}})(m^2_1 + m^2_2 + m^2_3 + m^2_4 - 3s), 
\label{eq44}\\
\lambda_1^{I=1} &=& 0.
\end{eqnarray}
Similarly, taking the interactions of $B^*_s \bar{B}^*_s \to B^*_s \bar{B}^*_s$ and $B^* \bar{B}^* \to B^*_s \bar{B}^*_s$, which now require $\phi$ and $K^*$ exchange, we also get
\begin{eqnarray}
\lambda_{1s}^{I=0} &=&  \frac12 \, g^2 \frac{1}{m^2_{\phi}} (m^2_1 + m^2_2 + m^2_3 + m^2_4 - 3s), \\
\lambda_{1s}^{I=1} &=& 0, \\
\lambda_{1m}^{I=0} &=&  \frac{1}{\sqrt{2}} \, g^2 \frac{1}{m^2_{K^*}} (m^2_1 + m^2_2 + m^2_3 + m^2_4 - 3s), \\
\lambda_{1m}^{I=1} &=& 0, 
\label{eq49}
\end{eqnarray}
This is a peculiar finding of the hidden gauge approach, which gives a null interaction in the $I=1$ sector. This would be in contradiction with the finding of the $Z_b(10610)$ and $Z_b(10650)$ resonances which appear very close to the $B \bar{B}^*$ and $B^* \bar{B}^*$ thresholds, such that, assuming they are molecular states, it implies that the interaction is weak or subdominant, as we are finding. In models where $\lambda_0, ~\lambda_1$ are fitted to some data, as in \cite{HidalgoDuque:2012pq}, one can get $I=1$ bound states.

Therefore, keeping HQSS constraints and determining LEC's by the local hidden gauge approach, finally  Eqs. \eqref{eq:ji00}$-$\eqref{eq:ji20} can be simplified and combined as follow,
\begin{itemize}
\item $I=0$, $J=0$
\[
\left. \phantom{(}
\begin{array}{cc}
\phantom{\text {$\lambda_{1m}$ } } &
\phantom{\text {$\lambda_{1m}$ } } \\
B \bar{B} &  B_s \bar{B}_s 
\end{array}
\right. \phantom{)_{ I=0,J=0}}
\]
\begin{equation}
\left(
\begin{array}{cc}
  \text{ $\lambda_{1}$ } &
   \text{ $\lambda_{1m}$ } \\ \\
 \text {$\lambda_{1m}$ } &
  \text{ $\lambda_{1s}$ }  \\
\end{array}
\right)_{ I=0,J=0}
\label{eq50}
\end{equation}

\item $I=0$, $J=0,1,2$
\[
\left. \phantom{(}
\begin{array}{cc}
\phantom{\text {$\lambda_{1m}$ } } &
\phantom{\text {$\lambda_{1m}$ } } \\
B^* \bar{B}^* &  B_s^* \bar{B}_s^* 
\end{array}
\right. \phantom{)_{ I=0,J=0,1,2}}
\]
\begin{equation}
\left(
\begin{array}{cc}
  \text{ $\lambda_{1}$ } &
   \text{ $\lambda_{1m}$ } \\ \\
 \text {$\lambda_{1m}$ } &
  \text{ $\lambda_{1s}$ }  \\
\end{array}
\right)_{ I=0,J=0,1,2}
\end{equation}

\item $I=0$, $J=1$
\[
\left. \phantom{(}
\begin{array}{cc}
\phantom{\text {$\lambda_{1m}$ } } &
\phantom{\text {$\lambda_{1m}$ } } \\
B \bar{B}^* &  B_s \bar{B}_s^* 
\end{array}
\right. \phantom{)_{ I=0,J=1}^{ C=-}}
\]
\begin{equation}
\left(
\begin{array}{cc}
  \text{ $\lambda_{1}$ } &
   \text{ $\lambda_{1m}$ } \\ \\
 \text {$\lambda_{1m}$ } &
  \text{ $\lambda_{1s}$ }  \\
\end{array}
\right)_{ I=0,J=1}^{ C=-}
\label{eq52}
\end{equation}

\[
\left. \phantom{(}
\begin{array}{cc}
\phantom{\text {$\lambda_{1m}$ } } &
\phantom{\text {$\lambda_{1m}$ } } \\
B \bar{B}^* &  B_s \bar{B}_s^* 
\end{array}
\right. \phantom{)_{ I=0,J=1}^{ C=+}}
\]
\begin{equation}
\left(
\begin{array}{cc}
  \text{ $\lambda_{1}$ } &
   \text{ $\lambda_{1m}$ } \\ \\
 \text {$\lambda_{1m}$ } &
  \text{ $\lambda_{1s}$ }  \\
\end{array}
\right)_{ I=0,J=1}^{ C=+}
\label{eq53}
\end{equation}

\end{itemize}

\section{The coupled channel approach for the heavy quark sector}
\label{secicca}

The scattering matrix is evaluated by solving the coupled channels Bethe-Salpeter equation in the on shell factorization approach of \cite{angels,ollerulf}
\begin{equation}
T = [1 - V \, G]^{-1}\, V,
\label{eq:Bethe}
\end{equation}
where the kernel $V$ has been discussed in the former section and the propagator $G$ is the loop function of two mesons, which is given by
\begin{equation}
G(s) = i \int\frac{d^{4}q}{(2\pi)^{4}}\frac{1}{(P-q)^{2}-m^2_1+i\varepsilon}\,\frac{1}{q^{2}-m^2_2+i\varepsilon},
\label{eq:G}
\end{equation}
where $m_1, ~m_2$ are the masses of the mesons, $q$ is the four-momentum of one meson, and $P$ is the total four-momentum of the systems, thus, $s=P^2$. The integral for the $G$ function, Eq. \eqref{eq:G}, is logarithmically divergent. There are two methods to regularize it. One is the dimensional regularization and the analytic expression can be seen in \cite{ollerulf} with a scale $\mu$ and the subtraction constant $a(\mu)$ as free parameter, 
\begin{eqnarray}
G(s) &=&\frac{1}{16\pi^2}\big\{a_{\mu}+\textmd{ln}\frac{m^2_1}{\mu^{2}}+\frac{m^2_2-m^2_1+s}{2s}\textmd{ln}\frac{m^2_2}{m^2_1} \nonumber \\
&&+\frac{q_{cm}}{\sqrt{s}}\big[\textmd{ln}(s-(m^2_1-m^2_2)+2q_{cm}\sqrt{s})+\textmd{ln}(s+(m^2_1-m^2_2)+2q_{cm}\sqrt{s}) \nonumber \\
&&-\textmd{ln}(-s-(m^2_1-m^2_2)+2q_{cm}\sqrt{s})-\textmd{ln}(-s+(m^2_1-m^2_2)+2q_{cm}\sqrt{s})\big]\big\}\ ,
\label{eq:Gdr}
\end{eqnarray}
where $q_{cm}$ is the three-momentum of the intermediate mesons in the center mass frame. The other method to regularize it is using a cut-off momentum, performing the integration
\begin{equation}
G(s) = \int_0^{q_{max}} \frac{d^3 \vec{q}}{(2\pi)^{3}}\frac{\omega_1+\omega_2}{2\omega_1\omega_2}\,\frac{1}{P^{0\,2}-(\omega_1+\omega_2)^2+i\varepsilon},
\label{eq:Gco}
\end{equation}
where $\omega_i = \sqrt{\vec{q}\,^2+m_i^2},~(i =1\,2)$, and $q_{max}$ is the cut-off of the three-momentum, the free parameter. Also the analytic formula of Eq. \eqref{eq:Gco} can be seen in \cite{Guo:2005wp,Oller:1998hw}.

Normally at low energies, the two regularization methods are compatible and there are relationships between these free parameters, $a(\mu)$, $\mu$ and $q_{max}$ \cite{ollerulf} (see also Eq. (52) of \cite{GarciaRecio:2010ki}). At higher energies, as discussed in \cite{xiaooset,wuzou}, there are large differences even not far away from threshold. This is because, for heavy mesons one can accommodate large momentum transfers with small energy, and cut offs of reasonable range, of the order of $500-1000\mev$, already produce distorted $G$ functions at excitation energies of the order of $100\mev$. On the other hand, for energies below threshold the cut off method always gives $G$ negative, while the dimensional regularization can produce $G$ positive, leading to an unwanted result of bound states with repulsive potentials (see discussion in \cite{xiaooset}).

We are interested in bound states, so, we will rely upon the cut off method. Then it is useful to recall how the on shell Bethe-Salpeter equation that we use here are derived with a Quantum Mechanical approach. This is done in \cite{gamer} and one can write,
\begin{equation}
V(\vec{q}\,',\vec{q}\,) = \langle \vec{q}\,'|\hat{V}|\vec{q}\, \rangle \equiv v\, f(\vec{q}\,') f(\vec{q}\,).
\label{eq:Vv}
\end{equation}
Then one shows in \cite{gamer} that the $T$ matrix factorizes like Eq. \eqref{eq:Vv} and one has
\begin{equation}
T(\vec{q},\vec{q}\,') = \langle \vec{q}\,|\hat{T}|\vec{q}\,' \rangle \equiv t\, f(\vec{q}\,) f(\vec{q}\,'),
\label{eq:Tt}
\end{equation}
and then the Lippmann-Schwinger equation becomes
\begin{equation}
t = [1 - v \, G]^{-1}\, v,
\label{eq:Bethe2}
\end{equation}
but now
\begin{equation}
G(s) = \int \frac{d^3 \vec{q}}{(2\pi)^{3}} f^2(\vec{q}\,) \frac{\omega_1+\omega_2}{2\,\omega_1\,\omega_2}\,\frac{1}{P^{0\,2}-(\omega_1+\omega_2)^2+i\varepsilon}.
\label{eq:Gco2}
\end{equation}
Once again we can write the integral equation as an algebraic equation \cite{angels}. Note that Eq. \eqref{eq:Bethe2} has the same format as Eq. \eqref{eq:Bethe}, but, the matrices $t,~v$ are defined by Eqs. \eqref{eq:Vv} and \eqref{eq:Tt}, and the loop function $G(s)$ is changed to Eq. \eqref{eq:Gco2} which absorbs a momentum dependent form factor from the factorized potential. A form factor $f(\vec{q}\,)$ that appears in our approach is discussed in \cite{xiaooset} and comes from the light vector meson exchange. It is obtained from the vector meson propagator keeping the three momentum exchange, ignoring the energy exchange. Hence,
\begin{equation}
f(\vec{q}\,) = \frac{m_V^2}{\vec{q}\,^2+m_V^2}. \label{eq:formf}
\end{equation}
This allows us to keep for $v$ the same potentials that have used before, and the effect of the form factor of Eq. \eqref{eq:formf} is absorbed in the $G$ function, Eq. \eqref{eq:Gco2}, that now becomes convergent. Certainly, for dynamical reasons, or just to account for missing channels in the approach, one still has some freedom in $q_{max}$ which we shall use in the results section.

Note that the introduction of $q_{max}$ in the $G$ function, is equivalent to multiplying the form factor $f(\vec q)$ by the step function $\theta(q_{max} - q)$. The potential in coordinate space can be obtained by Fourier transforming Eq. (\ref{eq:Vv}). A smaller cut off would imply a wider potential and a more spread wave function. The spread of the wave function would be relevant when discussing the variation of the couplings as one changes the cut off (see also \cite{gamer}).

The molecules appear as the poles of the $t$ matrix given in Eq. (\ref{eq:Bethe2}). The coupling of a given resonance
of mass $m_R$ to the $i^{th}$ channel can be obtained through:
\begin{equation}
g_i^2 = \lim_{s \rightarrow m_r^2} (s-m_R^2) t_{ii}
\end{equation}
Instead of taking this limit, which would require a high precision determination of $m_R$, the limit can be expressed as a loop integral in the complex $s$ plane:
\begin{equation}
g_i^2 = \frac{1}{2 \pi i} \oint t_{ii}ds 
\label{eq60}
\end{equation}
where the integral is over a closed path in the complex $s$ plane around the pole at $s=m_R^2$ and not crossing the branch cuts.

\section{Results and discussion}

In our formalism we used the following values for the masses of the mesons: $m_B=5.28$ GeV, $m_{B_s}=5.37$ GeV, $m_{B^*}=5.325$ GeV, $m_{B_s^*}=5.415$ GeV, 
$m_{K^*} = 0.892$ GeV, $m_\omega=0.783$ GeV, $m_\rho=0.775$ GeV, $m_\phi=1.019$ GeV. The leptonic decay constant of the pion is taken as $f_\pi = 93$ MeV. 
%Since the mass of $\Upsilon$ meson is much larger than the $K^*$, $\rho$, $\omega$ or $\phi$ masses, diagrams in which $\Upsilon$ is exchanged are neglected.

An important parameter in the analysis is the cutoff used to regularize the loop functions. 
In the heavy quark limit, the value of $q_{max}$ should be independent of quark flavor. To see this, note that in the heavy quark limit,
the binding energy of states is independent of the heavy quark mass. In terms of the $t$ matrix, this means that the position of
the poles of the $t$ matrix with respect to the threshold should be independent of the heavy quark mass, i.e., $vG$ should scale as $m_Q^0$ in the heavy quark limit.
Since the potential $v$ scales as $m_Q^2$ in the heavy quark limit\footnote{When the $v$ potential in this field theoretical approach is converted into the ordinary potential in Quantum Mechanics, the latter becomes of the order
$m_Q^0$ as shown in \cite{juanxiao}.}, the $G$ function should scale as $m_Q^{-2}$ to cancel the $m_Q$ dependence
of the potential $v$.
In the definition of the $G$ function, Eq. (\ref{eq:Gco}), if one makes the following approximations:
\begin{gather}
w_i \simeq m_Q, \nonumber \\
\frac{1}{(P^{0})^2 - (w_1+w_2)^2 + i \epsilon} \simeq \frac{1}{4B m_Q},
\end{gather}
where $B$  is the binding energy, the $G$ function can be estimated as
\begin{eqnarray}
G \simeq \frac{1}{4 B m_Q^2} \int_{q < q_{max}} \frac{d^3 q}{(2 \pi)^3} f(q^2)^2.
\end{eqnarray}

Since $G$ has to scale as $m_Q^{-2}$ in the heavy quark limit, the integral has to scale as $m_Q^0$, and hence
$q_{max}$ should be flavor independent in the heavy quark limit.

Requiring that the
presented formalism predicts a bound state of mass $3720$ MeV, as found in \cite{dany}, when the $B$ meson masses are replaced by the analogous
$D$ meson masses, yields the value $q_{max} = 415$ MeV. Assuming that this cutoff is independent of the heavy flavor, 
the same value is used in the $B$-meson sector. To estimate the errors due to variation of this cutoff, the spectrum is also
analyzed using twice this value: $q_{max} = 830$ MeV. These values are also consistent with the typical scales proposed in \cite{juanmanolo}.

A similar analysis of the $G$ function obtained using dimensional regularization, would yield the $m_Q$ dependence
of the subtraction constant $a_\mu$. In the dimensional regularization, there appear two constants $\mu$ and $a_\mu$.
These two constants are not independent and any change in one of them can be compensated by a change in the other one.
Using this freedom, we will fix $\mu=(m_1+m_2)/2$, i.e. the average of the masses of particles running in the loop.
Although any other choice of $\mu$ would be equally valid, this choice leads to a simple heavy quark limit of $a_\mu$, and at the
same time keeps the symmetry of the integral defining the $G$ function under the exchange of the masses $m_1 \leftrightarrow m_2$.

In the expression of the $G$ function given in Eq. (\ref{eq:Gdr}), substituting $s_R = (m_1+m_2-B)^2$, $m_i = m_Q + \lambda_i$, and expanding around
$1/m_Q =0$, the scaling of the $G$ function, as a function of the heavy quark mass at the resonance is obtained as:
\begin{eqnarray}
G =&& \frac{1}{16 \pi^2} \left[ a_\mu + \pi \sqrt{\frac{B}{m_Q}} - 2 \frac{B}{m_Q} \right.
\nonumber \\
&&\left. +\frac{1}{8}\pi \left(\frac{1}{m_Q}\right)^{3/2}
   \sqrt{B} (2\lambda_1+2\lambda_2-3B) + {\cal O}\left(\frac{1}{m_Q^2}\right) 
   \right]\label{heavylimitofdr}
\end{eqnarray}
As one can see, it is not easy to make $G$ scale as $m_Q^{-2}$ in a wide range of energies. Actually, problems arising from the use of the standard dimensional regularization formula in the $B$ meson sector were already discussed in \cite{wuzou}. There the $G$ function in dimensional regularization was made to match at the threshold the cut off result. This guaranteed that the energy dependence below threshold was very similar in both methods. But this similarity holds only in a very narrow region of about $10$ MeV. Beyond this region, both below and above, the results were quite different. These problems and the fact that we could match the $m_Q^{-2}$ behavior of $G$ in the cut of method makes this latter choice preferable and this is what we follow in this paper.

In principle, by exchanging $K^*$ mesons, the hidden strange sector is coupled to the non-strange one. 
When discussing coupled channels with such a small cut off for coupled channels with different masses, some technical details are in order. If we study bound states of $B\bar B$ and we add the $B_s \bar B_s$ channel,
the $G$ function for $B_s \bar B_s$ in the energy region around the $B \bar B$ threshold is negligible and
then the effect of the $B_s \bar B_s$ coupled channel is washed away.

The reverse has technical problems. If we investigate a possible state around the $B_s \bar B_s$ threshold,
separated by $180$ MeV from the $B \bar B$ threshold, the $B \bar B$ state would have a momentum of $980$ MeV/c, which is bigger than
the cut-off chosen. This means that the $G$ function for $B \bar B$ around the $B_s \bar B_s$ threshold will be unrealistic
with the small cut-off chosen and we can not use this method. It is better to argue that the $B \bar B$ state will not influence any
possible $B_s \bar B_s$ bound state in the same was as the $B_s \bar  B_s$ did not influence the bound state $B \bar B$. The only difference
is that the $B_s \bar B_s$ bound state could decay to the $B \bar B$ channel, but the disconnection of these states will make its width also small.
We can even estimate this width by taking $\mbox{Re} G_{B \bar B}=0$ around the $B_s \bar B_s$ threshold which we expect on physical grounds, but keeping
$\mbox{Im} G_{B \bar B}$ which can be calculated analytically to be:
\begin{eqnarray}
\mbox{Im} G = - \frac{1}{8 \pi} \frac{q_{cm}}{\sqrt s} f(q_{cm}^2)^2
\end{eqnarray}
where $q_{cm}$ is defined after Eq. (\ref{eq:Gdr}). 

In all the cases analyzed, coupled channels wash away the second pole, which is dominantly a hidden strange state. This second pole has a weak strength even in the single channel case.
The origin  of the lack of a second pole in the coupled channel case can  be traced back to the potential $v$. When the effects of coupled channel analysis are taken into account,  the dominant contribution to the determinant of the potential is proportional to\footnote{Here, we are assuming that $m_\rho = m_\omega$} $m_{K^*}^4 - m_\phi^2 m_\rho^2$ with a small correction from the mass difference of the hidden strange sector and the non-strange one (see Eqs.(\ref{eq44})$-$(\ref{eq49})). Since $m_{K^*}^2 \simeq m_\phi m_\rho$, this determinant is very small, hence one of the eigenvalues is very close to zero. This means that, in the corresponding  channel, which is mostly hidden strange state, the mesons do not interact and hence can not form a bound state.

In the $J^{PC}=2^{++}$ sector, the available channels are the $B^* \bar B^*$ and $B_s^* \bar B_s^*$. When the coupled channel effects are taken into account, the $t$ matrix has a single pole which is located at $m_R=10613$ MeV ($m_R=10469$ MeV) when $q_{max}=415$ MeV ($830$ MeV).
This corresponds to a binding energy of  $37$ MeV ($181$ MeV) with respect to the  $B^*\bar B^*$ threshold.
In Table \ref{table2pp} we present the masses and the couplings of this resonance to various channels for $q_{max}=415$ MeV and $q_{max}=830$ MeV. It is observed that both the binding energy and the couplings strongly depend on the value of the cut off chosen. Increasing the cut off from $q_{max}=415$ MeV to $q_{max}=830$ MeV changes the binding energy by about $140$ MeV whereas the couplings increase by a factor of two. This increase in the couplings is expected since as one increases the cut off, the potential has larger extent in momentum space, and hence the wave functions become narrower in coordinate space. A narrower $S$ wave wave function necessarily has a larger value at the origin. Since the couplings are proportional to the wave function at the origin, as the wave function gets narrower its value at the origin increases and the coupling grows.

As mentioned before, we deem the value of the lower cut off more realistic, and also in tune with \cite{juanmanolo}. The value obtained with $q_{max} = 830$ MeV should be considered a generous upper bound.

In Table \ref{table2ppn}, we present the properties of resonances if the coupled channel effects are ignored.
It is seen that the properties of the lighter resonance changes slightly by the removal of the coupled channel effects. Its
mass increases by $3$ MeV ($31$ MeV)  and its coupling to
the $B^*\bar B^*$ state is reduced by about $5\%$ ($10\%$) if the cut off is taken as $q_{max}=415$ Mev ($830$ MeV).
Since the coupled channel effects are ignored, this resonance does not couple to $B_s^*\bar B_s^*$. In this case, a weakly bound second pole is also observed in the $B^*_s\bar B_s^*$ channel. This second pole has a binding energy of $2$ MeV ($18$ MeV) when the cut off is taken to be $q_{max}=415$ MeV ($830$ MeV). This binding energy is more than ten times smaller than the binding of the lighter resonance. The coupling of this heavier resonance to $B^*_s\bar B^*_s$ is also about four times smaller that the coupling of the lighter resonance to the $B^*\bar B^*$ channel.

\begin{table}[ht]
     \renewcommand{\arraystretch}{1.5}
     \setlength{\tabcolsep}{0.4cm}
\centering
\caption{The couplings to various channels for the poles in the $J^{PC} = 2^{++}$ channel for $q_{max}=415$ MeV (left panel) and 
$q_{max}=830$ MeV (right panel)}
\begin{tabular}{ccc| |ccc}
\hline\hline
$10613$ & $B^* \bar B^*$ & $B_s^* \bar B_s^*$ & $10469$ & $B^* \bar B^*$ & $B_s^* \bar B_s^*$  \\
\hline
$g_i$ &$18703$ & 9955 & $g_i$ & $38112$ & $20290$\\
\hline
\end{tabular}
\label{table2pp}
\end{table}

\begin{table}[ht]
     \renewcommand{\arraystretch}{1.5}
     \setlength{\tabcolsep}{0.4cm}
\centering
\caption{The couplings to various channels for the poles in the $J^{PC} = 2^{++}$ channel for $q_{max}=415$ MeV (left panel) and 
$q_{max}=830$ MeV (right panel) ignoring coupled channels}
\begin{tabular}{ccc| |ccc}
\hline\hline
$10616$ & $B^* \bar B^*$ & $B_s^* \bar B_s^*$ & $10500$ & $B^* \bar B^*$ & $B_s^* \bar B_s^*$  \\
\hline
$g_i$ &$17708$ & $0$ & $g_i$ & $34719$ & $0$\\
\hline
$10828$ & $B^* \bar B^*$ & $B_s^* \bar B_s^*$ & $10812$ & $B^* \bar B^*$ & $B_s^* \bar B_s^*$ \\
\hline
$g_i$ &  $0$ & $4252$ & $g_i$ & $0$ & $9484$ \\
\hline
\end{tabular}
\label{table2ppn}
\end{table}

In a  $B^*\bar B^*$ molecule, the vector mesons can also combine in a total spin $1$ ($J^{PC}=1^{+-}$) or $0$ ($J^{PC}=0^{++}$) state. In
the heavy quark limit considered in this work, these channels are degenerate with the total spin $2$ state. Hence, their properties are identical to the properties of resonances shown in Tables \ref{table2pp} and \ref{table2ppn}.

Besides the $J=1$ combination in the $B_{(s)}^* \bar B^*_{(s)}$ channel, there are four other states with $J=1$ formed by $B_{(s)} \bar B_{(s)}^*$. These states have quantum number $J^{PC}=1^{++}$ and $J^{PC}=1^{+-}$ and are $B_{(s)} \bar B^*_{(s)} -$ c.c. and $B_{(s)} \bar B_{(s)}^* +$ c.c. respectively. These channels are degenerate in the heavy quark limit (see Eqs. (\ref{eq52}) and (\ref{eq53})). The properties of resonances in these channels are shown in Tables \ref{table1pm} and \ref{table1pmn}.
As in the $J^{PC}=2^{++}$ channel, if coupled channels are taken into account, there is only one resonance. This resonance has a binding energy of $37$ MeV ($180$ MeV) with respect to the $B \bar B^*$ threshold. Compared with the previous case, the binding energy is found to be degenerate with the binding energies obtained in the $J^{PC}=2^{++}$ channel. Due to the smaller mass of the $B\bar B^*$ system compared to the $B^*\bar B^*$,  the couplings to various channels are slightly smaller.\footnote{The couplings are related to the wave function at the origin. Due to the smaller masses, the wave function spreads more, reducing the value of the wave function at the origin.} The results obtained when coupled channel effects are ignored are presented in Table \ref{table1pmn}. As in the case when coupled channel effects are taken into account, the binding energies are degenerate with the corresponding case in $J^{PC}=2^{++}$ sector, and the couplings are slightly reduced.

\begin{table}[ht]
     \renewcommand{\arraystretch}{1.5}
     \setlength{\tabcolsep}{0.4cm}
\centering
\caption{The couplings to various channels for the poles in the $J^{PC} = 1^{+-}$ and $J^{PC}=1^{++}$ channels for $q_{max}=415$ MeV (left panel) and 
$q_{max}=830$ MeV (right panel)}
\begin{tabular}{ccc| |ccc}
\hline\hline
$10568$ & $B \bar B^*\pm$c.c. & $B_s \bar B_s^*\pm$c.c. & $10425$ & $B \bar B^*\pm$c.c. & $B_s \bar B_s^*\pm$c.c.  \\
\hline
$g_i$ &$18583$ & 9910 & $g_i$ & 37867 & 20199 \\
\hline
\end{tabular}
\label{table1pm}
\end{table}
\begin{table}[ht]
     \renewcommand{\arraystretch}{1.5}
     \setlength{\tabcolsep}{0.4cm}
\centering
\caption{The couplings to various channels for the poles in the $J^{PC} = 1^{+-}$  and $J^{PC}=1^{++}$ channels for $q_{max}=415$ MeV (left panel) and 
$q_{max}=830$ MeV (right panel) ignoring coupled channels}
\begin{tabular}{ccc| |ccc}
\hline\hline
$10571$ & $B \bar B^*\pm$c.c. & $B_s \bar B_s^*\pm$c.c. & $10455$ & $B \bar B^*\pm$c.c. & $B_s \bar B_s^*\pm$c.c.  \\
\hline
$g_i$ &$17591$ & $0$ & $g_i$ & $34485$ & $0$ \\
\hline
$10782.9$ & $B\bar B^*\pm$c.c. & $B_s \bar B_s^*\pm$c.c. & $10768$ & $B \bar B^*\pm$c.c. & $B_s \bar B_s^*\pm$c.c. \\
\hline
$g_i$ & $0$  & $4223$ & $g_i$  & $0$ & $9433$ \\
\hline
\end{tabular}
\label{table1pmn}
\end{table}

\begin{table}[ht]
     \renewcommand{\arraystretch}{1.5}
     \setlength{\tabcolsep}{0.4cm}
\centering
\caption{The couplings to various channels for the poles in the $J^{PC} = 0^{++}$ channel for $q_{max}=415$ MeV (left panel) and 
$q_{max}=830$ MeV (right panel)}
\begin{tabular}{ccc| |ccc}
\hline\hline
$10523$ & $B \bar B$. & $B_s \bar B_s$ & $10380$ & $B \bar B$ & $B_s \bar B_s$  \\
\hline
$g_i$ &$18538$ & 9865 & $g_i$ & $37760$ & $20102$ \\
\hline
\end{tabular}
\label{table0pp}
\end{table}

\begin{table}[ht]
     \renewcommand{\arraystretch}{1.5}
     \setlength{\tabcolsep}{0.4cm}
\centering
\caption{The couplings to various channels for the poles in the $J^{PC} = 0^{++}$ channel for $q_{max}=415$ MeV (left panel) and 
$q_{max}=830$ MeV (right panel) ignoring coupled channel effects}
\begin{tabular}{ccc| |ccc}
\hline\hline
$10526$ & $B \bar B$. & $B_s \bar B_s$ & $10410$ & $B \bar B$ & $B_s \bar B_s$  \\
\hline
$g_i$ &$17551$ & $0$ & $g_i$ & $34401$ & $0$ \\
\hline
$10738$ & $B \bar B$ & $B_s \bar B_s$ & $10723$ & $B \bar B$ & $B_s \bar B_s$ \\
\hline
$g_i$ &  $0$ & $4195$ & $g_i$ & $0$ & $9381$\\
\hline
\end{tabular}
\label{table0ppn}
\end{table}

In Tables \ref{table0pp} and \ref{table0ppn}, we present our results for the $J^{PC}=0^{++}$ sector. In this sector, the new states are $B\bar B$ and $B_{s} \bar B_s$. The similarities that we observed when comparing the $J^{PC}=2^{++}$ and $J^{PC}=1^{+\pm}$ sectors also exists when the $J^{PC}=0^{++}$ channel is compared with the previous cases, i.e. the binding energies are degenerate with the previous cases and due to the even smaller total mass in the $B_{(s)}\bar B_{(s)}$, the couplings are smaller. The binding obtained here for $B \bar{B}$ with the small cut off is very similar to the one obtained in \citep{Li:2012mqa} using the extended chiral quark model, where vector mesons are allowed to be exchanged between quarks, with clear similarities with the dynamics of the local hidden gauge approach.

Note that the observed degeneracies are consistent with the results obtained in \cite{HidalgoDuque:2012pq,Nieves:2012tt,Hidalgo-Duque:2013pva}. In these works, it is shown that in the heavy quark limit, in the spectrum of molecules of $\bar Q q$ and $\bar q Q$ where $Q$ is a heavy quark and $q$ is any other quark, the $J^{PC}=0^{++}$, $1^{++}$, $2^{++}$ and $1^{+-}$ states have degenerate binding energies. Furthermore, there are two other states with $J^{PC}=0^{++}$ and $1^{+-}$ that have degenerate binding energies which are not necessarily degenerate with the previous four. In our work, we observe that all six states have degenerate binding energies. These degeneracies can be observed from Eqs. (\ref{eq:ji00})$-$(\ref{eq:ji20}). If one uses the HQSS basis, Eqs. (\ref{eq:ji00}) and (\ref{eq:ji101}) will be brought to a block diagonal form. One of these blocks in both of these matrices is identical to the matrix given in Eq.  (\ref{eq:ji102}), resulting in four degenerate states. The other block is given by
\begin{equation}
\left(
\begin{array}{cc}
  \text{ $\lambda_{0}$ } &
   \text{ $\lambda_{0m}$ } \\ \\
 \text {$\lambda_{0m}$ } &
  \text{ $\lambda_{0s}$ }  \\
\end{array}
\right)_{ I=0}
\label{eq:block}
\end{equation}
and this block in each of the two matrices correspond to the other two degenerate states. In the approximations used in this work, the low energy constants $\lambda_0$ and $\lambda_1$ are equal (see discussions leading to Eqs. \eqref{eq:degeneracy1}$-$\eqref{eq:degeneracy}). Then automatically we get the uncoupled potentials of Eqs. \eqref{eq50}$-$\eqref{eq53}.
Hence, in this approximation, all six states should be degenerate in binding, and this is indeed observed in our analysis.

\section{Conclusions}

In the present work we investigate the hidden beauty sector by combining the dynamics of the local hidden gauge Lagrangians extrapolated to $SU(4)$ with the constraints of Heavy Quark Spin Symmetry.
The $SU(4)$ symmetry is broken to $SU(3)$ symmetry by taking large masses for the mesons containing a $b$-quark.

It is shown that in the $I=1$ sector, the interaction is too weak in the current approach to form any bound states. In the $I=0$ sector, both the hidden strangeness and non-strange channels are analyzed. The results show that the binding energies in all the possible $J^{PC}$ channels are degenerate.

When the couplings between the hidden strange and non strange sectors are ignored, bound states are observed in both sectors. Hence there are a total of $6$ hidden beauty resonances, with binding energies $34$ MeV ($178$ MeV) with respect to the non strange threshold, and $6$ hidden beauty-hidden strange resonances, with binding energies $2$ MeV ($18$ MeV) with respect to the hidden strange threshold, for a cut off value of $q_{max}=415$ MeV ($q_{max}=830$ MeV). The hidden beauty-hidden strange resonances are found to be weakly bound.

Our prediction of the existence of resonances close to the the hidden strange threshold is not robust with respect to the effects of the coupled channels. When the coupled channel effects are taken into account they disappear, whereas the masses of the other resonances are only slightly modified. Hence we predict with confidence the existence of (at least) $6$ resonances in the hidden beauty sector, with possible other $6$  heavier resonances which are mainly hidden beauty-hidden strange resonances. 

The couplings of each resonance to the various channels are also analyzed and for the lighter resonance in each channel, the couplings are shown to depend very slightly on the couple channel effects. It is also shown that the couplings are quite sensitive to the value of the cut off used, hence they should be taken more as an order of magnitude estimate rather than precise predictions.
When any of these states is experimentally found, the tuning of the cut off to the observed energies will also allow to be more precise on the value of these couplings.

\section*{Acknowledgments}  

We would like to thank J. Nieves for much help and useful discussions.
This work is partly supported by the Spanish Ministerio de Economia y Competitividad and European FEDER funds under the contract number FIS2011-28853-C02-01, and the Generalitat Valenciana in the program Prometeo, 2009/090. We acknowledge the support of the European Community-Research Infrastructure Integrating Activity Study of Strongly Interacting Matter (acronym Hadron Physics 3, Grant Agreement n. 283286) under the Seventh Framework Programme of EU. One of us, A.O. , acknowledges support from TUBITAK under project number 111T706.

\end{document}